\documentclass[journal]{IEEEtran}
\ifCLASSINFOpdf
\else
\fi
\usepackage{graphicx}
\usepackage{xcolor}
\usepackage{graphics}
\usepackage{blindtext}
\usepackage{amsmath}
\usepackage{adjustbox}
\usepackage{amsfonts}
\usepackage[]{units}
\usepackage{siunitx}
\usepackage{pdfpages}
\usepackage{amssymb}
\usepackage{fixltx2e}
\usepackage{etoolbox}
\usepackage[noadjust]{cite}
\usepackage{algorithm}
\usepackage{algorithmic}

\usepackage{caption} 
\makeatletter
\def\do#1{\patchcmd{#1}{\thepage}{\null}{}{\GenericWarning{}{Could not patch \string#1}}}
\docsvlist{\@oddhead,\@evenhead,\ps@headings,\ps@IEEEtitlepagestyle,\ps@IEEEpeerreviewcoverpagestyle}
\makeatother

\newcommand*\olinea[1]{%
  \vbox{%
    \hrule height 0.5pt
    \kern0.3ex
    \hbox{%
      \kern -0.05em
      \ifmmode#1\else\ensuremath{#1}\fi
      \kern 0.1em
    }
  }
}

\newcommand*\olineb[1]{%
  \vbox{%
    \hrule height 0.5pt
    \kern0.3ex
    \hbox{%
      \kern -0.05em
      \ifmmode#1\else\ensuremath{#1}\fi
      \kern 0em
    }
  }
}

\begin{document}

\bstctlcite{Bibliografia:BSTcontrol}

\title{Centralized Power Control in Cognitive Radio Networks Using Modulation and Coding Classification Feedback}

\author{Anestis Tsakmalis, \IEEEmembership{Student Member, IEEE,} Symeon Chatzinotas, \IEEEmembership{Senior Member, IEEE,}\\and Bj\"{o}rn Ottersten, \IEEEmembership{Fellow, IEEE}
\thanks{This work was supported by the National Research Fund, Luxembourg under the CORE projects "SeMIGod: SpEctrum Management and Interference mitiGation in cOgnitive raDio satellite networks" and "SATSENT: SATellite SEnsor NeTworks for spectrum monitoring".}
\thanks{The authors are with the Interdisciplinary Centre for Security, Reliability
and Trust (SnT), University of Luxembourg. Email:\{anestis.tsakmalis, symeon.chatzinotas, bjorn.ottersten\}@uni.lu.}}


\maketitle


\begin{abstract}

In this paper, a centralized Power Control (PC) scheme and an interference channel learning method are jointly tackled to allow a Cognitive Radio Network (CRN) access to the frequency band of a Primary User (PU) operating based on an Adaptive Coding and Modulation (ACM) protocol. The learning process enabler is a cooperative Modulation and Coding Classification (MCC) technique which estimates the Modulation and Coding scheme (MCS) of the PU. Due to the lack of cooperation between the PU and the CRN, the CRN exploits this multilevel MCC sensing feedback as implicit channel state information (CSI) of the PU link in order to constantly monitor the impact of the aggregated interference it causes. In this paper, an algorithm is developed for maximizing the CRN throughput (the PC optimization objective) and simultaneously learning how to mitigate PU interference (the optimization problem constraint) by using only the MCC information. Ideal approaches for this problem setting with high convergence rate are the cutting plane methods (CPM). Here, we focus on the analytic center cutting plane method (ACCPM) and the center of gravity cutting plane method (CGCPM) whose effectiveness in the proposed simultaneous PC and interference channel learning algorithm is demonstrated through numerical simulations.

\end{abstract}



\IEEEpeerreviewmaketitle

\section{Introduction}

\IEEEPARstart{W}{}ithin the last years, wireless communications have faced a steadily growing demand of multimedia and other bandwidth consuming interactive services. Taking also into account the static assignment of the frequency bands, spectrum has reached a saturation point. Measurements of the spectrum usage though have shown that even if some segments are congested, most of them are being underutilized. This indicates that the static assignment scheme is inefficient and a dynamic architecture should be adopted. Towards this direction, the research community proposed a concept called Dynamic Spectrum Access (DSA) \cite{biban36}, which suggests that services not fully utilizing their assigned frequency band can coexist with others. The first step of this evolution is to retain the costly infrastructure and spectrum access protocols of some services operating in their assigned bands and implement flexible and intelligent radio devices with DSA abilities which will detect access opportunities in these bands and exploit them to serve their own service demands. This kind of radio is called in literature Cognitive Radio (CR) and is able to sense, understand, adapt and interact with its surroundings based on the user's demands and the environment's limitations \cite{biban21}.

A main function of the CRs is Spectrum Sensing (SS). Like any intelligent entity, the CR must first observe its environment in order to learn from it and then interact with it. The first SS approaches were mainly focused on the classic binary hypothesis testing of PU existence. Another way of enhancing the CR's senses is signal classification. This radio must be able not only to detect whether a PU signal exists but also to identify its kind and an interesting approach is to recognize the modulation and coding scheme (MCC) of the PU signal \cite{biban57, biban71}. As far as the modulation classification is concerned, features like the signal Higher Order Statistical (HOS) cumulants which have distinctive theoretical values among different modulation schemes \cite{biban25} are estimated and then fed into a powerful classification tool, the Support Vector Machine (SVM) \cite{biban29}. For the coding identification part, the exploited statistical features are the log-likelihood ratios (LLRs) of the received symbol samples \cite{biban38, biban39}. The detection technique in this case involves the comparison of the average LLRs of the error syndromes derived from the parity-check relations of each code.

Other crucial functions of the cognition cycle of the CR are the learning and interacting procedures. In this paper, the latter abilities concern the transmit power of the unlicensed cognitive users, also called Secondary Users (SUs), which coexist in the same frequency band with the PUs and they are described as PC. One major category of cognitive PC techniques accomplishing this coexistence is the underlay one \cite{biban82}. In the underlay CR scenarios, on which we focus here, SUs may transmit in the PU frequency bands as long as the induced to the PU interference is under a certain limit. Therefore, the CRN should learn how to manage properly the transmit powers of its users. As mentioned before, the first stage of the DSA evolution will be the deployment of CRs (SUs) capable of using their acute senses in order to access frequency bands already used by older communication technologies (PUs), also referred to as legacy systems. Therefore, the transmit power strategy under which the SUs will access the frequency band of the PUs cannot rely on an access protocol that cooperates with the PUs' one to enter the frequency band, simply because the PUs' infrastructure or protocols cannot be easily changed. A practical approach for the CRN would be the SUs to be coordinated by a CBS using a dedicated control channel, which signifies a centralized PC scheme \cite{biban89}. Still, the CRN must acquire some kind of knowledge about the CR-to-PU channel gains and hence the induced interference to the PU.

Since no cooperation between the PU and SU systems is expected, accurate Channel State Information (CSI) about the interference channels cannot be obtained. In the CR context though, a common approach is the CR individual user or network to exploit a PU link state feedback, monitor how this changes because of the CRN operation and thus estimate the CR-to-PU channel gains. In previous work, this was extracted from the binary ACK/NACK feedback of the reverse PU link \cite{biban50, biban52, biban72, biban73} for PC or beamforming purposes. Here, we must mention that acquiring this binary feedback would require the implementation of the complete PU receiver on the CR side to decode the PU message and retrieve its ACK/NACK feedback. In addition to the hardware complexity issue, this rises security issues for the exploitation of the PU message. Also, to decode the PU message the sensed PU signal on the CR side must have a minimum required SINR, which might not always be the case.

\subsection{Contributions}

In this paper, a centralized PC method aided by interference channel gain estimation is demonstrated which concerns a PU and multiple SUs and maximizes the total SU throughput subject to maintaining the PU QoS. This case study considers the PU link changing its MCS based on an ACM protocol and operating in its assigned band together with a CRN accessing this band and \textit{having knowledge of this ACM protocol}. Our idea is to detect the PU MCS in a cooperative way in the CBS which gathers the sensed MCC feedback from all the SUs through a control channel and combines them using a hard decision fusion rule and subsequently to exploit this multilevel feedback, instead of the hard to obtain binary ACK/NACK packet, in order to learn the CR-to-PU channel gains. This channel knowledge is acquired by having the SUs constantly changing their transmit power under the CBS instructions and checking whether the CRN caused the PU MCS to change, a clearly probing procedure. Furthermore, a novel technique is developed so that the probing/learning method can be performed concurrently with the pursuit of the CRN maximum throughput and without this affecting the learning convergence time.

The mathematical formulation of this scenario is basically an optimization problem, the maximization of the total SU throughput, under an unknown inequality constraint, the preservation of the aggregated interference below a threshold to maintain the PU MCS. In this paper, reaching the optimization objective and learning the unknown constraint by using the MCC feedback are performed in parallel. Ideal learning approaches for this problem setting are the CPMs, whose high learning rate is not affected severely by the sampling procedure, the CRN power allocation. In this case, the sampling procedure is choosing training data (the SU transmit power levels) which satisfy the optimization objective subject to the until that learning step estimated interference constraint. Here, we focus on two of the fastest CPMs, the ACCPM and the CGCPM. The ACCPM has been used by the research community for enhancing the speed of various learning methods and the CGCPM has attracted attention mostly due to its theoretically fastest convergence rate.

This design novelty of exploiting the MCC feedback and combining a learning procedure with an optimization problem in such a way delivers specifically the following contributions:

\begin{itemize}

  \item For the first time, the MCS degradation is used as a multilevel feedback of the induced interference. The complexity of the MCC module is much simpler than that of an actual decoder which is used in underlay CR scenarios of other papers to obtain the ACK/NACK packets of the PU link. In addition, the MCC feedback provides more information than the binary feedback and therefore improves the learning rate of the interference constraint.
  \item A simple cooperative MCC procedure is introduced based on plurality voting.
  \item A PC mechanism for static interference channels is proposed where maximizing the total SU throughput subject to an unknown PU interference constraint is taking place simultaneously with an interference channel gain learning process. This mechanism is an enhanced variation of the scheme proposed in \cite{biban73}.
  \item A dynamic adaptation of this mechanism is proposed for slow fading channels which is taking into account a window of the most recently observed feedback.
  \item Simulations show a convergence rate for the CPM based methods faster than the one of the benchmark method developed in \cite{biban80} and furthermore a learning speed superiority of the CGCPM based method compared to the ACCPM based technique \cite{biban73}.

\end{itemize}

\subsection{Structure and Notation} 

The remainder of this paper is structured as follows: Section II reviews in detail
prior work related to cognitive scenarios using a PU link feedback. Section III provides the system model and the problem formulation. Section IV analyzes the simultaneous PC and interference channel learning algorithm. Section V shows the simulation results obtained from the application of the proposed techniques and compares them with a benchmark method. Finally, Section VI gives the concluding remarks and future work in this topic.


\section{Related work}

Previous work in the field of cognitive underlay PC has considered a great variety of assumptions, protocols, system models, optimization variables, objective functions, constraints and other known or unknown parameters. The general form of the underlay CR scenarios is the optimization of a SU system metric, such as total throughput, worst user throughput or SINR of every SU, subject to QoS constraints for PUs, like SINR, data rate or outage probability \cite{biban82}. Moreover, the research community has formed combinations of the aforementioned PC problems with beamforming patterns, base station assignment, bandwidth or channel allocation and time schedules, which led to more complicated joint problems, but with the same basic form. Based on the coordination or cooperation of the CR network, PC is separated in two categories, the centralized and the decentralized.

The most important issue arising from cognitive scenarios is the knowledge of the interference channel gains. In prior work, this piece of information was either assumed known \cite{biban70} or within some uncertainty limits \cite{biban83, biban64}. Although, this presumption helped to devise sophisticated optimization problems, it is not applicable in most cases. Here, we describe scenarios with one common characteristic, no prior knowledge of the CR transmitter to PU receiver channel gain. This assumes that a learning mechanism of the interference channel gains is implemented by a central decision maker or each SU individually. A necessary condition for the learning process is the availability of a feedback which is usually acquired by a SS technique, assuming no cooperation between the CRN and the PU system. An interesting idea was proposed in \cite{biban81} called proactive SS, where the SU probes the PU and senses its effect from the PU power fluctuation. Also, the exploitation of the MCC feedback, used in our work, is suggested briefly by the authors of \cite{biban81} in a footnote. Primarily though, the most common piece of information being used to estimate the interference channel gains is the binary feedback, which is often obtained by eavesdropping the PU feedback channel and detecting the ACK/NACK packet.

In the decentralized or distributed underlay scenarios, the binary feedback has been used to enable CRs apply Reinforcement Learning procedures, like Q-Learning and Bush-Mosteller Learning, to regulate the aggregated interference to the PU \cite{biban53} and additionally reach a throughput optimization objective \cite{biban52}. Formulating this problem as a repeated PC game and employing Game Theory to analyse it \cite{biban52} has been a critical contribution to explain the behaviour of such a system and prove the convergence of decentralized learning methods. Also, pricing distributed PC schemes have been developed under outage probability constraints \cite{biban50}.

As far as the centralized underlay research work is concerned, a central decision maker, the CBS, must learn the interference channel gains, elaborate an intelligent selection of the operational parameters of the SUs, such as their transmit power, and communicate it to them. Even though, distributed PC underlay scenarios have been investigated thoroughly, the centralized PC problem combined with interference channel gain learning is still an unexplored area. Notably, the most sophisticated and fast methods suitable for the CBS learning the interference channel gains of multiple SUs with the use of feedback come from multiple antenna underlay cognitive scenarios. In this point, we need to explain how channel learning in beamforming problems can easily be translated as channel learning in centralized PC problems. If you assume that each one of the multiple antennas corresponds to a SU in a CRN, then coordinating the beamforming vectors in order to estimate the CR to PU channel gains is no different than a CBS coordinating the transmit powers of a CRN for the same purpose. In fact, designing the transmit powers is actually much simpler than composing each antenna's complex coefficient in the beamforming scenarios, since in PC no phase parameters are incorporated.

Previous researchers in this field have exploited slow stochastic approximation algorithms \cite{biban79, biban84}, the one-bit null space learning algorithm (OBNSLA) \cite{biban72} and an ACCPM based learning algorithm \cite{biban78}. The last two approaches were introduced as channel correlation matrix learning methods with the ACCPM based technique outperforming the OBNSLA. All these learning techniques are based on a simple iterative scheme of probing the PU system and getting a feedback indicating how the PU operation is changed. One other thing in common of the aforementioned work is the discrimination of the channel learning phase and the transmission phase which is optimum to an objective, like the maximum total throughput or maximum SINR transmission. Thus, the optimization objective is achieved only after the learning process is terminated. Nonetheless, the ideal would be to tackle them jointly and learn the interference channel gains while at the same time pursuing the optimization objective without that affecting the learning convergence time. On this rationale, the authors of \cite{biban73} proposed an ACCPM based learning algorithm where probing the PU system targets to both learning channel correlation matrices and maximizing the SNR at the SU receiver side. In this paper, we exploit this idea in the underlay PC problem by using the MCC sensing feedback instead of the binary ACK/NACK packet captured from the PU feedback channel. In this problem formulation, learning the interference channel gains from each SU to the PU receiver is performed concurrently with maximizing the total SU throughput under an interference constraint which depends on these channel gains. Additionally, remarks are made on this method, enhancements are introduced and its results are compared to a benchmark learning technique \cite{biban80}.


\section{System Model and Problem Formulation}

\begin{figure}[ht!]
\centering
\includegraphics[scale=0.65, trim=10 0 0 10]{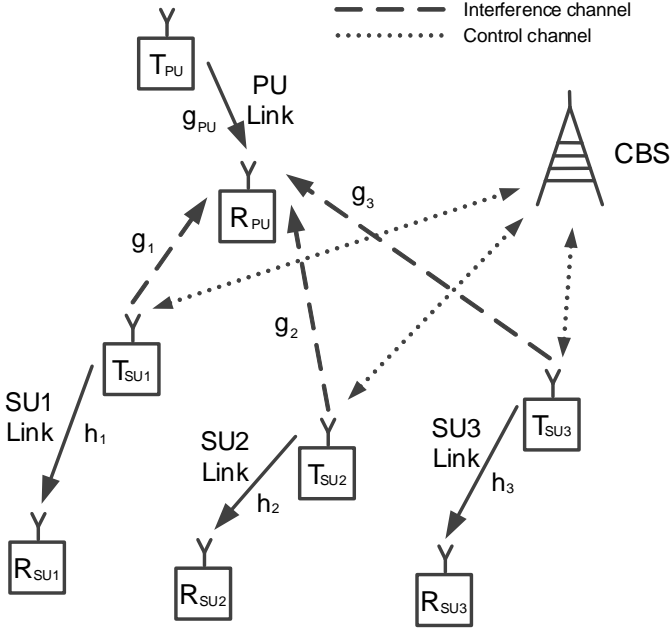}
\caption{The PU system and the CR network}
\label{fig1}
\end{figure}

Consider a PU link and $N$ SU links existing in the same frequency band as shown in Fig. \ref{fig1}. Furthermore, a multiple access method (e.g. FDMA) allows SU links to operate in the PU frequency band and not to interfere with each other. The examined scenarios in this paper are considering the PU channel gain to be static and the unknown interference channel gains static and slow fading. Here we focus on channel power gains $g$, which in general are defined as $g=\|c\|^{2}$, where $c$ is the complex channel gain. From this point, we will refer to channel power gains as channel gains.

As far as the interference to the PU link is concerned, this is caused by the transmitter part of each SU link to the receiver of the PU link. Taking into account that the SU links transmit solely in the PU frequency band, the aggregated interference on the PU side is defined as:

\begin{equation}
I_{PU}=\mathbf{g^\intercal}\mathbf{p}
\label{eq1}
\end{equation}
where $\mathbf{g}$ is the interference channel gain vector $[g_{1},...,g_{N}]$ with $g_{i}$ being the SU\textsubscript{i}-to-PU interference channel gain and $\mathbf{p}$ is the SU power vector $[p_{1},...,p_{N}]$ with $p_{i}$ being the SU\textsubscript{i} transmit power. Additionally, the SINR of the PU is defined as:

\begin{equation}
SINR_{PU}=10\log\left(\frac{g_{_{PU}}p_{_{PU}}}{I_{PU}+N_{PU}}\right)\mbox{dB}
\label{eq2}
\end{equation}
where $g_{_{PU}}$ is the PU link channel gain, $p_{_{PU}}$ is the PU transmit power and $N_{PU}$ is the PU receiver noise power.

In this paper, we address the problem of total SU throughput ($U_{SU}^{tot}$) maximization without causing harmful interference to the PU system, which can be written as:
\begin{subequations}
 \label{eq3:optim}
 \begin{align}
 & \underset{\mathbf{p}}{\text{maximize}}
 & & U_{SU}^{tot}(\mathbf{p})=\sum\limits_{i=1}^{N}\log\left(1+\frac{h_{i}p_{i}}{N_{i}}\right) \label{eq3:a} \\
 & \text{subject to}
 & & \mathbf{g^\intercal}\mathbf{p} \leq I_{th} \label{eq3:b} \\
 &
 & & \mathbf{0} \leq \mathbf{p} \leq \mathbf{p_{max}} \label{eq3:c}
 \end{align}
\end{subequations}
where $\mathbf{p_{max}}=[p_{max_{1}},...,p_{max_{N}}]$ with $p_{max_{i}}$ being the maximum transmit power level of the SU\textsubscript{i} transmitter, $h_{i}$ is the channel gain of the SU\textsubscript{i} link and $N_{i}$ is the noise power level of the SU\textsubscript{i} receiver. The channel gain parameters $h_{i}$ and the noise power levels $N_{i}$ are considered to be known to the CRN and not changing in time. An observation necessary for tackling this problem is that the $g_{i}$ gains normalized to $I_{th}$ are adequate for defining the interference constraint. Therefore, the new version of \eqref{eq3:b} will be:

\begin{equation}
\mathbf{\tilde{g}^\intercal}\mathbf{p} \leq 1
\label{eq4}
\end{equation}
where $\mathbf{\tilde{g}}=\frac{\mathbf{g}}{I_{th}}$.

This optimization problem is convex and using the Karush-Kuhn-Tucker (KKT) approach a capped multilevel waterfilling (CMP) solution is obtained \cite{biban70} for each SU\textsubscript{i} of the closed form:
\begin{equation}
p^{*}_{i}=\left\{
  \begin{array}{cc}
   p_{max_{i}} & \mbox{if $\frac{1}{\lambda \tilde{g}_{i}}-\frac{N_{i}}{h_{i}} \geq p_{max_{i}}$}\\
   0 & \mbox{if $\frac{1}{\lambda \tilde{g}_{i}}-\frac{N_{i}}{h_{i}} \leq 0$}\\
   \frac{1}{\lambda \tilde{g}_{i}}-\frac{N_{i}}{h_{i}} & \mbox{otherwise}
  \end{array}, \; i = 1, \ldots, N
  \right.
\label{eq5}
\end{equation}
where $\lambda$ is the KKT multiplier of the interference constraint \eqref{eq4} and which can be determined as presented in \cite{biban70}.

\begin{figure}[ht!]
\centering
\includegraphics[scale=0.485, trim=0 -30 0 -10]{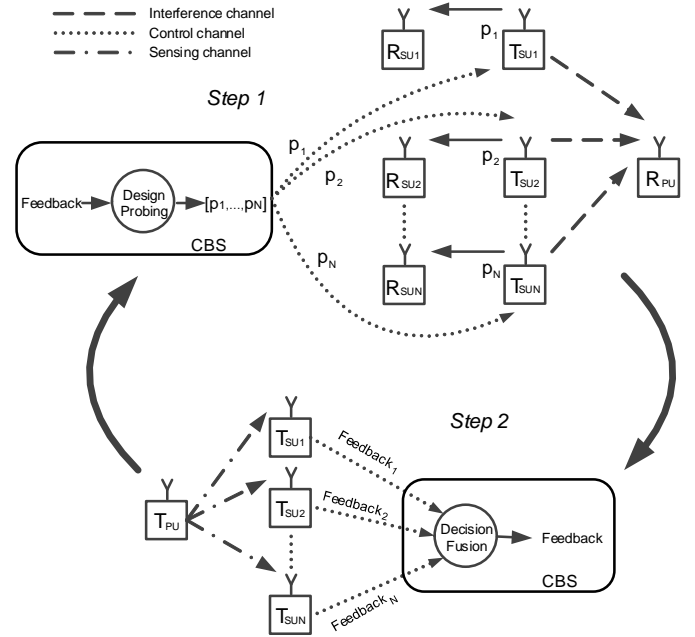}
\caption{The algorithm steps}
\label{fig12}
\end{figure}

Even though this problem setting is well known and already investigated, in the next sections we will demonstrate how to cope with it without knowing the interference constraint \eqref{eq4}. An algorithm will be described which combines learning the normalized interference channel gain vector $\mathbf{\tilde{g}}$ of \eqref{eq4} with the use of an implicit PU CSI feedback and maximizing $U_{SU}^{tot}$ without causing harmful interference to the PU system. As illustrated in Fig. \ref{fig12}, the basic recurrent steps of this algorithm will be:

\begin{description}
  \item[]
  \item[\textit{Step 1}: Design probing and probe the PU]
  \item[]
  \item[\textit{Step 2}: Sense feedback and infer the probing impact]
  \item[]
\end{description}

\subsection{The Multilevel Modulation and Coding Classification Feedback}

In this section, we deal with the MCC feedback, which is the enabler of the interference constraint learning defined by the unknown $\tilde{g}_{i}$ parameters. This is basically the \textit{Step 2} of the algorithm as mentioned previously. Initially, the outputs of the cooperative MCC procedure have to be noted. In our previous work \cite{biban80}, a cooperative MCC method is described where all the SUs are equipped with a secondary omnidirectional antenna only for sensing the PU signal and an MCC module which enables them to identify the MCS of the PU. Specifically, each SU collects PU signal samples, estimates the current MCS, forwards it through a control channel to the CBS and finally the CBS using a hard decision fusion rule combines all this information to get to a decision based on a plurality voting system. After casting every vote, the CBS identifies the PU MCS.

Even though plurality voting is a simple and not sophisticated method which elects the MCS
value that appears more often than all of the others, it produces the correct voting output under the condition that some SUs have sensing channels of moderate quality. Its equivalent voting system for binary data fusion, the majority one, has been used by the research community to improve the detection and false alarm probabilities with satisfactory results. Additionally, it is appropriate in multiple hypothesis tests where the statistics of the classification metric are not easy to handle, as in our case.

Taking into account strong interference links may have a severe effect on the MCS chosen by the PU link, which changes to more robust modulation constellations and coding rates depending on the level of the $SINR_{PU}$. Let $\{MCS_{1},..,MCS_{J}\}$ denote the set of the MCS candidates of the ACM protocol and $\{\gamma_{1},..,\gamma_{J}\}$ the corresponding minimum required $SINR_{PU}$ values, which whenever violated, an MCS adaptation happens. Furthermore, consider these sets arranged such that $\gamma$'s appear in an ascending order. Here, it has to be pointed out that it is reasonable to assume that the CRN has some a priori knowledge of the standard of the legacy PU system whose frequency band attempts to enter and therefore the CRN can be aware of the PU system ACM protocol and of its $\gamma_{j}$ values. Assuming that $N_{PU}$ and the received power remain the same at the PU receiver side, the $\{\gamma_{1},..,\gamma_{J}\}$ values correspond to particular maximum allowed $I_{PU}$ values, designated as $\{I_{th_{1}},..,I_{th_{J}}\}$. Hence, whenever the PU is active, for every $MCS_{j}$ it can be inferred that $I_{PU}$ lies within the interval $(I_{th_{j+1}},I_{th_{j}}]$, where $I_{th_{j}}$ is the interference threshold over which the PU is obliged to change its transmission scheme to a lower order modulation constellation or a lower code rate and $I_{th_{j+1}}$ is the interference lower limit below which the PU can change its transmission scheme to a higher order modulation constellation or a higher code rate. Still, the actual values of these thresholds are unknown to the CRN, since the CRN cannot be aware of the $N_{PU}$ and the received power at the PU receiver side.

This groundwork predisposes us how to transform the MCS feedback into a multilevel piece of information, instead of exploiting it as binary \cite{biban80}. Nevertheless, in our interference channel learning problem we have to encounter the fact that the CRN has no knowledge of $\{I_{th_{1}},..,I_{th_{J}}\}$. To this direction, the observation that learning the interference channel gain vector $\mathbf{g}$ is equivalent to learning the normalized interference channel gain vector $\mathbf{\tilde{g}}$ of \eqref{eq4} is essential. Now, taking as reference the PU MCS when the SU system is not transmitting at all, $MCS_{ref}=MCS_{k}$, and the corresponding $\gamma_{ref}=\gamma_{k}$, where $k\in \{1,..,J\}$, the following $\gamma$ ratios can be defined:

\begin{equation}
c_{j}=\frac{\gamma_{j}}{\gamma_{ref}}
\label{eq6}
\end{equation}
where $j\neq k$ and $j\in \{1,..,J\}$. Supposing a high $SNR_{PU}$ regime, $g_{_{PU}}p_{_{PU}}\gg N_{PU}$, the $I_{th_{j}}$ ratios can also be determined as:

\begin{equation}
\frac{I_{th_{j}}}{I_{th_{ref}}}=\frac{\gamma_{ref}}{\gamma_{j}}=\frac{1}{c_{j}}
\label{eq7}
\end{equation}
where $I_{th_{ref}}$ is the interference threshold of $MCS_{ref}$.

The knowledge of these ratios has a great significance for our normalization process. Let $MCS_{ref}$ be the sensed MCS when the CRN is silent and no probing occurs, $\mathbf{p}=\mathbf{0}$, and $MCS_{j}$ be the deteriorated MCS after the SU system probed the PU using an arbitrary SU power vector $\mathbf{p}$. The information gained by the CBS as mentioned before is that:

\begin{equation}
I_{th_{j+1}}<\mathbf{g^\intercal}\mathbf{p}\leq I_{th_{j}}
\label{eq8}.
\end{equation}
These inequalities can be rewritten using the $I_{th}$ ratios as:

\begin{equation}
\frac{I_{th_{ref}}}{c_{j+1}}<\mathbf{g^\intercal}\mathbf{p}\leq \frac{I_{th_{ref}}}{c_{j}}\Longleftrightarrow \frac{1}{c_{j+1}}<\mathbf{\tilde{g}^\intercal}\mathbf{p}\leq \frac{1}{c_{j}}
\label{eq9}
\end{equation}
where $\mathbf{g}$ is normalized like in \eqref{eq4} with $I_{th}=I_{th_{ref}}$ as $\mathbf{\tilde{g}}=\frac{\mathbf{g}}{I_{th_{ref}}}$.

Thus, when a probing procedure is applied to the PU system, the MCC feedback allows us to detect where the probing SU power vector lies within the feasible region more accurately without searching uselessly the power vector feasible region. The former inequalities \eqref{eq9} can also be formulated in a further normalized version:

\begin{equation}
  \begin{array}{cc}
   \mathbf{\tilde{g}^\intercal}\mathbf{\tilde{p}_{u}} > 1\\ \\
   \mathbf{\tilde{g}^\intercal}\mathbf{\tilde{p}_{l}} \leq 1
  \end{array}
\label{eq10}
\end{equation}
where $\mathbf{\tilde{p}_{l}}=c_{j+1}\mathbf{p}$ and $\mathbf{\tilde{p}_{u}}=c_{j}\mathbf{p}$. This advantage of using the multilevel MCC feedback instead of a simple binary indicator, such as the ACK/NACK packet of the PU link, will be employed by the learning technique described in the latter section in order to estimate the unknown interference channel gain vector, $\mathbf{\tilde{g}}$, and reach the optimization objective defined by \eqref{eq5}.
  

\section{The Simultaneous Power Control and Interference Channel Learning Algorithm}

The main problem tackled in this paper is to find a fast learning method aided by feedback and whose training samples can be chosen by an intervening process without that affecting the convergence time of the learning part. Essentially, this concerns the design of the probing which takes place in \textit{Step 1} of the algorithm as illustrated in Fig. \ref{fig12}. This idea was first explored as a cognitive beamforming problem by the authors of \cite{biban73} who managed by properly probing the PU system and using only ACK/NACK packets of the PU feedback channel to simultaneously learn channel correlation matrices and maximize the SNR at the SU receiver side by applying a CPM, the ACCPM. CPMs are iterative techniques which accumulate inequalities in a sequential way to localize a search point \cite{biban86}. These inequalities represent a convex uncertainty set in the search space where the point lies and which is cut in every time step using a CPM. The target is to gradually diminish this uncertainty set and reach the search point within an error limit. Basically, the CPMs are extensions of the 1-dimension bisection method to higher dimensions and the inequalities represent the cutting planes in these higher dimensions.

In each CPM iteration, two pieces of information are needed to define a cut:

\begin{description}
  \item[]
  \item[$\bullet$ the center of the convex uncertainty set]
  \item[]
  \item[$\bullet$ a hyperplane passing through this center]
  \item[]
\end{description}

In this point, we shall explain how this framework can be applied to our learning problem. The goal of this learning procedure is to estimate the parameter vector $\mathbf{\tilde{g}}$ of the interference constraint as represented in \eqref{eq4} using the SU system probing power vectors as training samples. In this probing procedure, the SU system has the freedom of intelligently choosing the training samples in order to learn and not just receive them from a teaching process. In Machine Learning, this kind of learning is called Active Learning, where the learner actually chooses training samples that are more informative so that he can reach the learning solution faster, with less training samples and with less processing. The learning speed, and thus the smaller number of probing power vectors, is an essential part of the suggested idea, because of two main reasons. The SU system must learn the interference constraint fast so that first it will not interfere the PU and reduce the PU QoS for a long time and secondly it can apply this learning method in a fading channel environment. Ideal Active Learning methods for this task are the newly introduced to this field CPMs, which have attractive convergence properties because of their geometric characteristics. Still, the CPMs that we have been chosen are used to localize points in a search space. For this purpose, a conceptual trick must be used which in Machine Learning literature was introduced by Vapnik \cite{biban29} and is called the "version space duality". According to that, points in the training sample or feature space are hyperplanes in the parameter or version space and vice versa. Hence, when a learning procedure tries to estimate the parameters of a hyperplane (the version) it actually tries to localize a point in the parameter or version space. In our problem, the feature space corresponds to the training sample space or the power vector space and the version space to the parameter $\mathbf{\tilde{g}}$ space, where the point being sought is the endpoint of the interference channel gain vector. Another fact worth being noted is that the inequalities obtained by feedbacks (the labels of our training) are meaningful also in the parameter $\mathbf{\tilde{g}}$ space since they are linear inequalities with respect to $\tilde{g}_{i}$'s.

Besides the fast convergence time of detecting a point, the main advantage of CPMs is that the training sample, $\mathbf{p}$ in this case, can be chosen based on any rationale without that affecting the decrease of the uncertainty region in the parameter $\mathbf{\tilde{g}}$ space. This rationale can be in our problem the solution of the optimization problem defined in \eqref{eq5}. Hence, approaching the actual endpoint of the parameter vector $\mathbf{\tilde{g}}$ can happen in parallel with maximizing the SU system throughout, the optimization objective. More specifically, at each learning step the CPM only dictates the center of the uncertainty set, an estimation of $\mathbf{\tilde{g}}$, and the hyperplane/cutting plane passing through this center, which is actually determined by $\mathbf{p}$, can be the solution of \eqref{eq3:optim}. Since the chosen cutting plane passes through it, the SU system power allocation vector is considered to satisfy the equality of the so far estimated interference constraint.

\subsection{Details of the CPM application to our problem}

In the CPM literature, there has been an extensive analysis of the different center definitions of a convex set based on which each CPM is differentiated from the others. This paper examines the CGCPM and the ACCPM and their corresponding centers, the center of gravity and the analytic center. These are regarded as the two types of center points "deeper" in a convex set and therefore efficient in dissecting the uncertainty set more evenly, a necessary condition for reaching fast to the sought point. Now, consider that the initial sensing MCC feedback by the CRN when no probing occurs, $\mathbf{p}(0)=\mathbf{0}$, is $MCS_{ref}$. Following $t$ probing attempts, the CBS has collected $t$ MCC pieces of feedback which correspond to $t$ pairs of inequalities:

\begin{equation}
  \begin{array}{cc}
   \mathbf{\tilde{g}^\intercal}\mathbf{\tilde{p}_{u}}(k) > 1\\ \\
   \mathbf{\tilde{g}^\intercal}\mathbf{\tilde{p}_{l}}(k) \leq 1
  \end{array}, \; k = 1, \ldots, t
\label{eq11}.
\end{equation}
The \eqref{eq11} inequalities are derived as described in the previous section in the form of \eqref{eq10} and additionally consider inequalities coming from probing power vectors which do not cause MCS deterioration. In order to keep a single notation in \eqref{eq11} even for power vectors not degrading the PU MCS, the first inequality does not hold and $\mathbf{\tilde{p}_{l}}$ is regarded equal to $\mathbf{p}$ in this special case. An additional constraint for the $\tilde{g}_{i}$ parameters is that $\tilde{g}_{i}$'s have to be positive as channel gains:

\begin{equation}
\tilde{g}_{i}\geq0 ,\;\;\; i = 1, \ldots, N
\label{eq12}
\end{equation}
The inequalities \eqref{eq11} and \eqref{eq12} define a convex polyhedron $\mathcal{P}_{t}$, the uncertainty set of the search problem:

\begin{equation}
\mathcal{P}_{t}=\left\{\mathbf{\tilde{g}}\mid\mathbf{\tilde{g}}\geq\mathbf{0}, \mathbf{\tilde{g}^\intercal}\mathbf{\tilde{p}_{u}}(k) > 1, \mathbf{\tilde{g}^\intercal}\mathbf{\tilde{p}_{l}}(k) \leq 1,k = 1, \ldots, t\right\}
\label{eq13}
\end{equation}

In the CGCPM, the center of gravity $CG$ of the convex polyhedron $\mathcal{P}_{t}$ is calculated in vector form as:

\begin{equation}
\mathbf{\tilde{g}_{CG}}(t)=\frac{\int_{\mathcal{P}_{t}}\mathbf{\tilde{g}}\;dV_{\tilde{g}}}{\int_{\mathcal{P}_{t}}dV_{\tilde{g}}}
\label{eq14}
\end{equation}
where $V_{\tilde{g}}$ represents volume in the parameter $\mathbf{\tilde{g}}$ space. The advantages of the CGCPM are that its convergence to the point in search is guaranteed and that the number of the uncertainty set cuts or inequalities needed are of $\mathcal{O}(N\log_{2}(N))$ complexity \cite{biban86}. This convergence rate is ensured by the fact that any cutting plane passing through the $CG$ reduces the polyhedron volume by at least $37\%$ at each step. The main disadvantage of using the $CG$ is its calculation, a computationally expensive integration process in multiple dimensions. A way of bypassing this issue is the randomization solution proposed by the author of \cite{biban94} which computes an approximation of the $CG$. The general idea is to generate many random sample points within $\mathcal{P}_{t}$ by taking a random walk, the so called \textit{Hit and Run} method, and average them to find the $CG$.

In the ACCPM, the analytic center $AC$ of the convex polyhedron $\mathcal{P}_{t}$ is calculated in vector form as:

\begin{align}
\mathbf{\tilde{g}_{AC}}(t)=&\arg\min\limits_{\mathbf{\tilde{g}}}\left(-\sum\limits_{k=1}^{t}\log(\mathbf{\tilde{g}^\intercal}\mathbf{\tilde{p}_{u}}(k)-1) \nonumber \right.\\ &\left.-\sum\limits_{k=1}^{t}\log(1-\mathbf{\tilde{g}^\intercal}\mathbf{\tilde{p}_{l}}(k))-\sum\limits_{i = 1}^{N}\log(\tilde{g}_{i})\right)
\label{eq15}.
\end{align}
Interior point methods can be used to efficiently solve the optimization problem described in \eqref{eq15} with a computational complexity of $\mathcal{O}(\sqrt{t})$ and estimate the $AC$ which makes this center a tractable choice for CPMs \cite{biban91}. Furthermore, an upper bound for the number of inequalities needed to approach the sought point has been evaluated to prove the convergence of the ACCPM which is of $\mathcal{O}(N^{2})$ complexity, also referred to as iteration complexity. The applicability of the ACCPM due to the computationally cheap calculation of the $AC$ has found numerous learning applications, like the fast training of SVMs and solving optimization problems involving nondifferentiable functions in the Operational Research field.

\subsection{The Necessity of Exploration}

\begin{figure}[ht!]
\centering
\includegraphics[scale=0.15, trim=940 10 900 0]{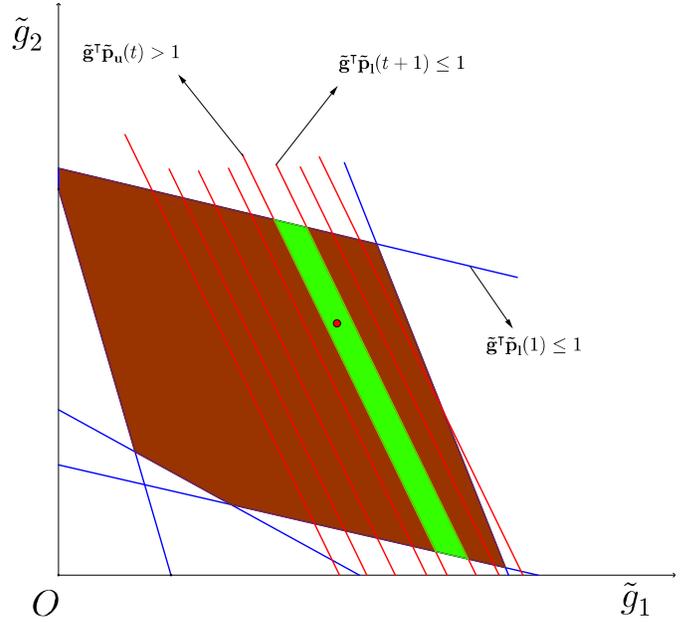}
\caption{The CPM in 2D when no exploration occurs}
\label{fig2}
\end{figure}

Even though this framework seems ideal for learning the interference constraint and at the same time pursuing the optimization objective, there is still a problem arising. The optimization part, which is responsible for choosing the training power vectors, focuses on cutting planes of specific direction as illustrated in Fig. \ref{fig2}. These training power vectors basically correspond to the power level ratios which maximize $U_{SU}^{tot}(\mathbf{p})$ and are subject to the \textit{initial} interference hyperplane estimation. Thus, they focus on specific power level ratios and contribute only in reducing uncertainty in this direction.

This indicates that choosing the training power vectors based solely on the optimization problem is not a good strategy. Instead, the SU system should start probing the PU system in an exploratory manner by diversifying initially the training power vectors and gradually, when enough knowledge of the interference constraint is obtained, shift to an exploitive behaviour which allocates power levels to the SUs specified by the optimization problem solution \eqref{eq5}.

The authors of \cite{biban73} proposed to make this shift from \textit{exploration} to \textit{exploitation} by mixing the optimization objective, the maximization of the SU received SNR, with a similarity metric of the beamforming vectors. The influence of this similarity metric in the design of these probing vectors was determined to be a decreasing function of time, so that the desirable transition could happen. This is a combination of two tactics known in the Machine Learning community as the $\epsilon$-decreasing and contextual-$\epsilon$-greedy strategies \cite{biban90} and according to which the choice of the training samples is performed using an exploration or else randomization factor, $\epsilon$. In these strategies, this factor decreases as time passes or depending on the similarity of the training samples, resulting in explorative behaviour at the beginning and exploitative behaviour at the end. Nevertheless, this logic not only requires tuning of the exploration factor time dependency according to performance results, but it also does not guarantee that enough diversification has occurred to reach the learning goal, which in the case of \cite{biban73} is the channel correlation matrix, since time on its own cannot be an indicator of approaching the exact values of the sought parameters.

The enhancement introduced in this paper is to relate the exploration factor, $\epsilon$, to the proximity of $\mathbf{\tilde{g}}(t)$ to $\mathbf{\tilde{g}}$, where $\mathbf{\tilde{g}}(t)=\mathbf{\tilde{g}_{CG}}(t)$ or $\mathbf{\tilde{g}}(t)=\mathbf{\tilde{g}_{AC}}(t)$ depending on the CPM. Clearly this depends on the geometry of $\mathcal{P}_{t}$, the region where we search. Towards this goal, a simple approximation of this convex polyhedron, the minimum bounding box containing it, is adopted. The minimum bounding box, $\mathcal{B}_{t}$, indicates how large the uncertainty region, $\mathcal{P}_{t}$, is and in order to compute this, we first need to solve the following $2N$ Linear Programs:

\begin{equation}
\tilde{g}_{max_{i}}(t)=\max\limits_{\tilde{g} \in \mathcal{P}_{t}} \tilde{g}_{i},i = 1, \ldots, N
\label{eq16}
\end{equation}
\begin{equation}
\tilde{g}_{min_{i}}(t)=\min\limits_{\tilde{g} \in \mathcal{P}_{t}} \tilde{g}_{i},i = 1, \ldots, N
\label{eq17}
\end{equation}
which provide us the boundaries for the values of $\tilde{g}_{i}$ at each step $t$. Now, let $\mathbf{V}(t)=\{\mathbf{v}_{1}(t),..,\mathbf{v}_{N_{v}}(t)\}$, where $N_{v}=2^{N}$, denote the set of the minimum bounding box vertices which are defined straightforward from the boundaries of $\tilde{g}_{i}$. A proximity metric of $\mathbf{\tilde{g}}(t)$ to $\mathbf{\tilde{g}}$ could be the euclidean distance of these points $d(\mathbf{\tilde{g}}(t),\mathbf{\tilde{g}})=\lVert \mathbf{\tilde{g}}(t)-\mathbf{\tilde{g}} \rVert$, but the problem is that $\mathbf{\tilde{g}}$ is unkonwn. To fix this, the proximity metric is chosen as the maximum distance of $\mathbf{\tilde{g}}(t)$ from a $\mathcal{B}_{t}$ vertex:

\begin{equation}
d_{max}(t)=\max\limits_{\mathbf{v}_{j}(t) \in \mathbf{V}(t)}d(\mathbf{\tilde{g}}(t),\mathbf{v}_{j}(t))
\label{eq18}
\end{equation}
which is an upper bound of $d(\mathbf{\tilde{g}}(t),\mathbf{\tilde{g}})$. The proposed error driven solution is to relate $\epsilon$ to this proximity metric, a variation of the tactic known as adaptive $\epsilon$-greedy strategy. According to this, the closer the learning algorithm gets to the exact value $\mathbf{\tilde{g}}$, the less exploration occurs and training power vectors are more relative to the optimization problem solution \eqref{eq5}. A simple design to adapt $\epsilon$ is:

\begin{equation}
\epsilon(t)=\left\{
  \begin{array}{cc}
   1-\frac{d_{th}}{d_{max}(t)} & \mbox{if $d_{max}(t) > d_{th}$}\\
   0 & \mbox{if $d_{max}(t) \leq d_{th}$}
  \end{array}
  \right.
\label{eq19}
\end{equation}
where the threshold $d_{th}$ is linked with the precision limit that the learning algorithm has. That signifies that once $d_{max}(t)$ passes below this threshold, the algorithm has reached the exact solution within an error bound and thus there is no need to explore, but to exploit and choose power vectors according to \eqref{eq5}.

Moreover, the usage of $\epsilon(t)$ has to be specified and the way the training power vectors are chosen in case of $\epsilon(t)>0$. As mentioned before, $\epsilon(t)$ is a randomization factor which imposes that the power vector must be chosen randomly with $\epsilon(t)$ probability and the reason for that is to differentiate the cutting hyperplanes passing through the AC or CG of the CPM procedure. This random selection of power vectors is better explained in the power vector space, the variable space. The random power vector has to satisfy first the equality version of the so far estimated interference constraint \eqref{eq4}:

\begin{equation}
\mathbf{\tilde{g}^\intercal}(t)\mathbf{p}=1
\label{eq20}
\end{equation}
and second the constraints \eqref{eq3:c}. Consequently, this random selection is translated into a uniform sampling on the simplex piece $\mathcal{S}(t)$ defined by \eqref{eq20} and \eqref{eq3:c}.

\subsection{The Static and Slow Fading Channel Formulation of the Algorithm}

To clarify all this process described thoroughly in the previous section, we present it in Algo. \ref{alg1}. Specifically, in the $t_{th}$ iteration of this process the CRN designs the probing vector $\mathbf{p}(t)$ and probes the PU system, which requires a $T_{p}$ period for the CBS to calculate and communicate $\mathbf{p}(t)$ to all SUs and for the CRN to actually probe the PU (\textit{Step 1} of Fig. \ref{fig12}), and the CBS detects the PU MCS, $MCS(t)$, which demands a $T_{s}$ period for all SUs to collect PU signal samples, extract their estimates of PU MCS, send them to the CBS and amass them to make the final MCS decision (\textit{Step 2} of Fig. \ref{fig12}).

\begin{algorithm}
\begin{algorithmic}
\STATE $t=0$
\STATE $\mathbf{p}(t)=\mathbf{0}$
\STATE Sense $MCS(t)$
\STATE Assume an initial $\mathbf{\tilde{g}}(t)$
\LOOP
\STATE $t=t+1$
\STATE Compute $\epsilon(t)$
\STATE Generate $rand \in (0,1)$ 
\IF{$rand\geq \epsilon(t)$}
\STATE Exploit: $\mathbf{\tilde{p}}(t)=\arg\max U_{SU}^{tot}$ s.t. $\mathbf{\tilde{g}^\intercal}(t)\mathbf{\tilde{p}}=1$
\ELSE
\STATE Explore: $\mathbf{\tilde{p}}(t)=$ random point $\in \mathcal{S}(t)$
\ENDIF
\STATE Sense $MCS(t)$
\STATE Create new pair of inequalities \eqref{eq11}
\STATE Compute $\mathbf{\tilde{g}}(t)$ using a CPM
\ENDLOOP
\end{algorithmic}
\caption{The Simultaneous Power Control and Interference Channel Learning Algorithm}\label{GA}
\label{alg1}
\end{algorithm}

A formulation for slow fading interference channels is also given with some modifications of Algo. \ref{alg1}. The solution proposed in this paper is window-based in contrast with the maximum likelihood concept suggested in \cite{biban73} which considered a probit modelling of each inequality age. To approach the case of slow fading interference channels, first we must take into account the grade of channel variation over time. For this purpose, a quasi static block fading modelling of the interference channels is chosen, according to which the interference channel gains remain constant within a block period, also called coherence time. Assuming that the coherence time $T_{c}$ of the interference channels is known and the same for all interference channels, the crucial problems we need to tackle is the asynchronous change of the interference channel gains and the lack of knowledge about the exact time an interference channel change occurs. In order to handle these issues, first we calculate how many probing and sensing time periods fit in the coherence time, approximately $t_{c}=\frac{T_{c}}{T_{p}+T_{s}}$. From these $t_{c}$ iteration periods which correspond to an equal number of probing power vectors and sensing inequality pairs, we recommend to use for the slow fading algorithm formulation the last $t_{w}=\lfloor\frac{t_{c}}{N}\rfloor$ inequality pairs to construct a time window from the $(t-t_{w})_{th}$ to the $t_{th}$ probing and sensing period. This actually changes the set of inequalities taken into account to compute the $\mathbf{\tilde{g}}(t)$ using a CPM in order to include only the latest $t_{w}$ inequality pairs:

\begin{equation}
  \begin{array}{cc}
   \mathbf{\tilde{g}^\intercal}\mathbf{\tilde{p}_{u}}(k) > 1\\ \\
   \mathbf{\tilde{g}^\intercal}\mathbf{\tilde{p}_{l}}(k) \leq 1
  \end{array}, \; k = t-t_{w}, \ldots, t
\label{eq21}.
\end{equation}
More precisely, the convex polyhedron in no longer defined by \eqref{eq11} and \eqref{eq12}, but by \eqref{eq21} and \eqref{eq12}.


\section{Results}

In this section, we provide simulation results to compare the performance of the benchmark method, shown in \cite{biban80}, and the CPM based methods proposed in this paper. The CPM based methods are an enhancement of the ACCPM based simultaneous channel correlation matrix learning and beamforming solution provided in \cite{biban73}. Furthermore, the CGCPM is tested to validate its theoretically faster convergence compared to that of the ACCPM. Additionally, the benefit of utilizing the multilevel MCC feedback instead of the binary ACK/NACK packet is demonstrated for all the aforementioned techniques. To prove the MCC feedback superiority, we have chosen the legacy PU system to be operating using an ACM protocol close to the outdated technical specifications of 802.11a/g with LDPC coding \cite{biban92, biban93}. The selected MCS set and the corresponding $\gamma$ values are:

\begin{table}[!h]
\caption{The PU ACM protocol}
\centering
\begin{tabular}{|| c || c ||}
\hline
$\mathbf{MCS}$ & $\boldsymbol{\gamma}$ \\ \hline
$BPSK\;1/2$ & $5\mbox{dB}$ \\
\hline
$BPSK\;3/4$ & $6\mbox{dB}$ \\
\hline
$QPSK\;1/2$ & $7\mbox{dB}$ \\
\hline
$QPSK\;3/4$ & $9\mbox{dB}$ \\
\hline
$16QAM\;1/2$ & $13\mbox{dB}$ \\
\hline
\end{tabular}
\label{table:1}
\end{table}
Also, the PU receiver is chosen to normally operate at $SINR_{PU}=20\mbox{dB}$ with no interference and $N_{PU}=-103\mbox{dBm}$ resulting to $MCS_{ref}=16QAM\;1/2$. The $I_{th}$ which corresponds to $16QAM\;1/2$, is unknown to the CRN and over which a PU MCS adaptation occurs, resulting to PU QoS deterioration, is $-97\mbox{dBm}$. Given the information in Table \ref{table:1}, the formulation of the $\gamma$ ratios can easily be written using \eqref{eq6} in order to construct the normalized inequality pairs \eqref{eq10}.

Another simulation parameter necessary to be defined is the selected threshold $d_{th}$ related to the precision limit of the learning algorithm and to the exploration factor design. This is chosen at $5 \%$ which signifies that once the learning error upper bound, $d_{max}(t)$, is below $5\%$ the algorithm no longer explores but solely exploits to achieve the CRN throughput maximization.

Initially, the static interference channel scenario is examined with N = 5 SUs which are dispersed uniformly within a $3\mbox{km}$ range around the PU receiver. The unknown to the CRN interference channel gains are assumed to follow an exponential path loss model $g_{i}=\frac{1}{d_{i}^{4}}$, where $d_{i}$ is the distance of the SU\textsubscript{i} from the PU receiver in metres. The last operational parameter concerning the SUs is their maximum transmit power, $p_{max_{i}}$, which is set to $23\mbox{dBm}$ for all SUs.

\begin{figure}[!h]
\centering
\includegraphics[scale=0.69, trim=25 10 0 10, clip]{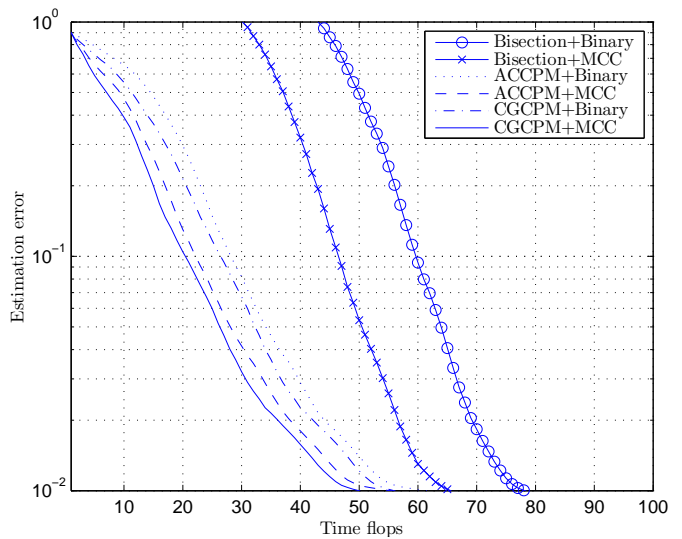}
\caption{Interference channel gain vector estimation error progress vs time of all method and feedback combinations for 5 SUs}
\label{fig3}
\end{figure}

Fig. \ref{fig3} shows the channel estimation error diagrams for the benchmark, ACCPM-based and CGCPM-based methods depending on the number of time flops where each time flop is the time period $T_{p}+T_{s}$ necessary to coordinate the CRN, probe the PU system, sense the MCC feedback and decide collectively the PU MCS. The interference channel gain vector estimation error metric at each time flop is defined as the normalized root-square error $\frac{\lVert \mathbf{\tilde{g}}(t)-\mathbf{\tilde{g}} \rVert}{\lVert \mathbf{\tilde{g}} \rVert}$. The error figure results are obtained as the average of the error metric defined earlier over $100$ SU random topologies, which deliver $100$ random draws of interference channel gain vectors $\mathbf{g}$. 

It can be clearly seen in Fig. \ref{fig3} that the CPM-based methods outperform the benchmark learning method. This occurs because the benchmark method may be the fastest Active Learning method in the training sample space, but the proposed CPM-based methods are performed in the version space, which appears to be more efficient. More specifically as far as the method comparison is concerned, for an estimation error approximately $1\%$, the benchmark method achieves convergence in $78$ and $65$ time flops for binary and MCC feedback respectively, whereas the corresponding numbers of time flops for the ACCPM-based technique are $61$ and $55$ and for the CGCPM-based one are $55$ and $50$. For the binary feedback, a gain of at least $17$ time flops is accomplished and for the MCC feedback the gain is at least $10$ time flops.  

Another outcome is that the utilization of the MCC feedback instead of the binary ACK/NACK packet reduces the convergence time significantly in the benchmark method and noticeably in the CPM-based learning methods. Specifically, for an estimation error of $1\%$, in the benchmark technique this gain of time flops is almost $13$ and in the CPM-based techniques it is nearly $6$. Even though the convergence time reduction is small in the CPM case, it is considered a notable enhancement considering that CPM-based techniques are already fast enough. The final conclusion derived from Fig. \ref{fig3} is about the comparison of the two CPM-based learning mechanisms. It is observed that the CGCPM-based scheme surpasses the ACCPM-based one and particularly for an estimation error of $1\%$ the CGCPM-based procedure outperforms the ACCPM-based one in the binary feedback case by $6$ time flops and in the MCC feedback case by $5$ time flops.

\begin{figure}[!h]
\centering
\includegraphics[scale=0.72, trim=38 10 0 10, clip]{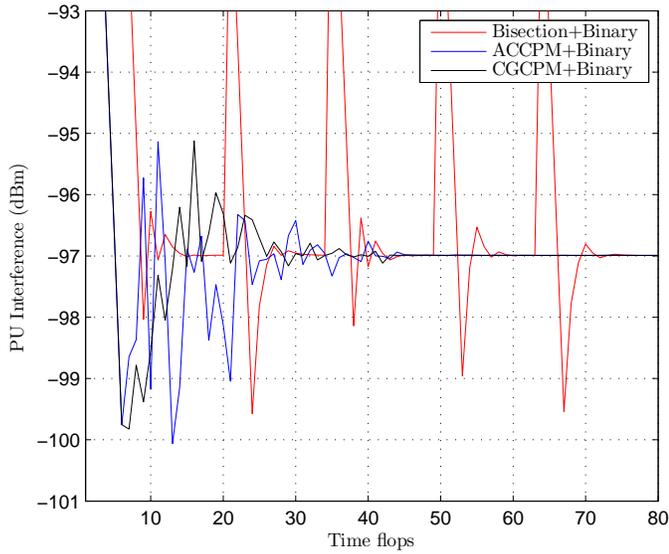}
\caption{$I_{PU}$ progress vs time using binary feedback}
\label{fig4}
\end{figure}

\begin{figure}[!h]
\centering
\includegraphics[scale=0.72, trim=38 10 0 10, clip]{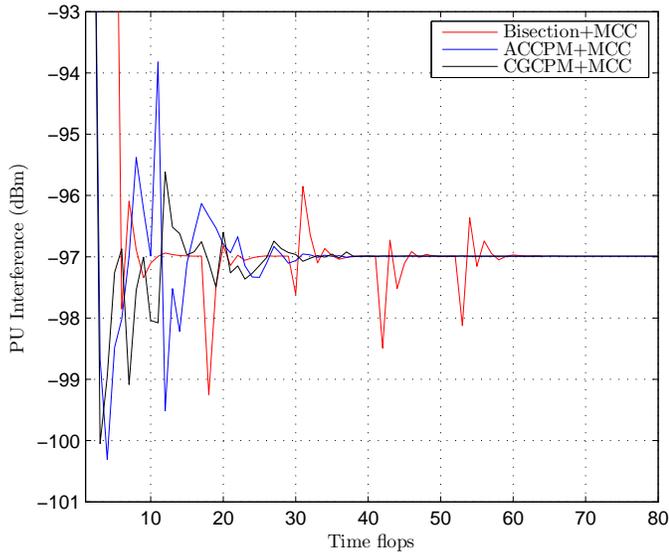}
\caption{$I_{PU}$ progress vs time using MCC feedback}
\label{fig5}
\end{figure}

The next diagrams show the aggregated interference caused to the PU during the simultaneous learning and CRN capacity maximization method for binary feedback in Fig. \ref{fig4} and for MCC feedback in Fig. \ref{fig5} of one random SU topology.
Originally, it is overt that taking advantage of the MCC feedback instead of the binary one causes smaller and fewer interference peaks and conduces to faster convergence. Secondly, it is observed that the CPM-based methods reach the learning objective faster than the benchmark method and finally that the CGCPM-based scheme converges to the PU interference threshold limit with less variations and more smoothly than the ACCPM-based.

\begin{figure}[!h]
\centering
\includegraphics[scale=0.72, trim=40 10 0 10, clip]{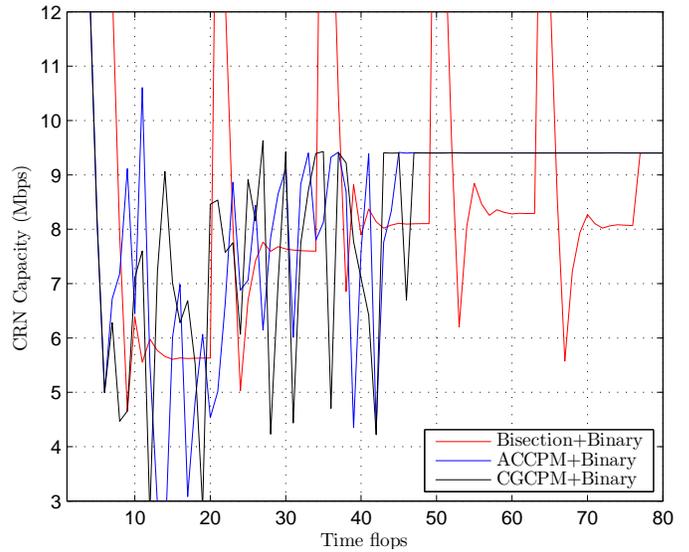}
\caption{CRN Capacity vs time using binary feedback}
\label{fig6}
\end{figure}

\begin{figure}[!h]
\centering
\includegraphics[scale=0.72, trim=40 10 0 10, clip]{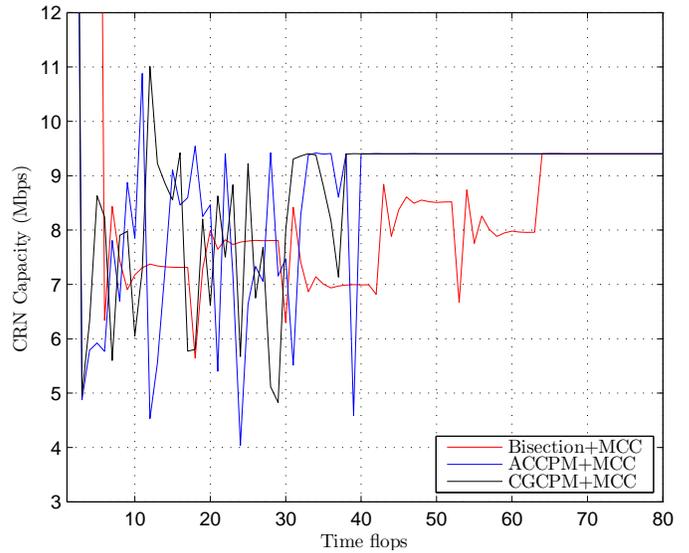}
\caption{CRN Capacity vs time using MCC feedback}
\label{fig7}
\end{figure}

The last diagrams of the 5 SU static scenario depict the CRN capacity progress vs time of the same SU topology for binary feedback in Fig. \ref{fig6} and for MCC feedback in Fig. \ref{fig7}.
The results of CRN capacity in Fig. \ref{fig6} and Fig. \ref{fig7} initially show, as stated before, the benefit of using the MCC feedback. The CRN capacity variations, which are mostly interpreted as CRN throughput degradation, are less in the MCC feedback scenarios. Comparing the methods, it has to be noted that the CGCPM case exhibits again a more graceful convergence to the maximum CRN capacity than that of the ACCPM case which is more obvious in Fig. \ref{fig7}.

\begin{figure}[!h]
\centering
\includegraphics[scale=0.69, trim=25 10 0 10, clip]{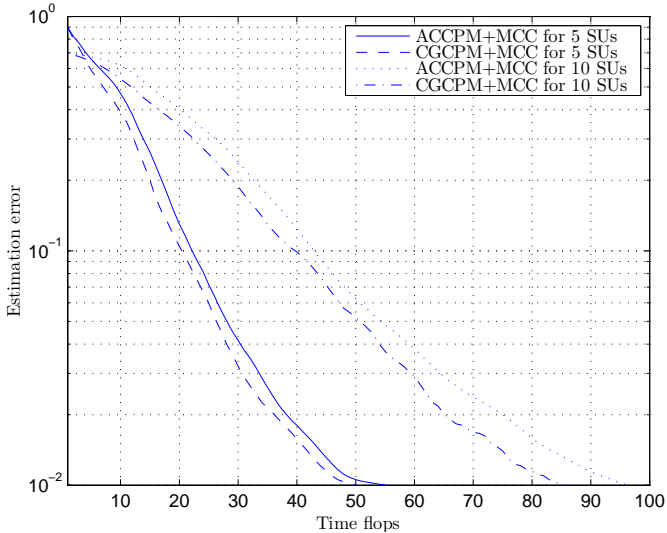}
\caption{Interference channel gain vector estimation error progress vs time of CPM based methods and MCC feedback for 5 and 10 SUs}
\label{fig11}
\end{figure}

To clearly show that the CGCPM based method is faster than the ACCPM based one, a fact indicated by CPM theory about their iteration complexities and mentioned in a previous section, we need to increase the problem dimensions, the number of the SUs. The next diagram in Fig. \ref{fig11} is about a static interference channel scenario with N = 10 SUs and exhibits the channel estimation error metric for the ACCPM-based and CGCPM-based methods with MCC feedback. Furthermore, the error performance of the same method and feedback combinations for N = 5 SUs are shown in the same diagram to prove that the convergence gain is increased as the number of the problem dimensions is increased.
As seen in Fig. \ref{fig11}, our variation of the ACCPM, which was used in \cite{biban73} to enhance the channel correlation matrix learning speed, achieves an estimation error $1\%$ at $95$ time flops, while the corresponding CGCPM based algorithm obtains the same error at $85$ time flops. This provides us a convergence gain of $10$ time flops which is increased compared to the 5 SU case and of course greater protection to the PU receiver with the CGCPM based method. Nevertheless, this gain in learning speed comes with a penalty. As noted in earlier section, the \textit{Hit and Run} calculation of the CGCPM requires the generation of many random samples within the polytope $\mathcal{P}_{t}$. The number of these samples grows exponentially with the number of problem dimensions. Hence, in order for the CBS, where the $CG$ computation takes place, to perform this calculation an exponentially increasing computational burden is needed. This means that the larger the CRN a CBS must coordinate, the more computations the CBS needs to perform in order to achieve the fastest convergence possible.

Subsequently, the proposed algorithms are tested for slow fading interference channels where $T_{c}$ is chosen to be equal to $250$ probing and sensing periods, $T_{p}+T_{s}$. The corresponding time window based on the empirical rule of $\lfloor\frac{t_{c}}{N}\rfloor$ for N = 5 SUs is $t_{w}=50$ inequality pairs and the rest of the algorithm settings remain the same with the fixed channel experiment case. In addition, $100$ random SU topology scenarios are generated for a duration of $3$ block periods which correspond to $750$ probing and sensing periods and where $2$ interference channel changes occur. Furthermore, it must be mentioned that in these experiments the benchmark method can be no longer used, since it can be only exploited for learning static interference channels, and that the binary feedback is not taken into account as it was proven earlier that it is inferior to the multilevel MCC feedback. Consequently, in this section we compare the performance of the CPM-based methods using the MCC feedback.

Once more, the first diagrams concern the learning error of the methods which depict an average of all the random SU topology simulations.

\begin{figure}[!h]
\centering
\includegraphics[scale=0.71, trim=29 10 0 10, clip]{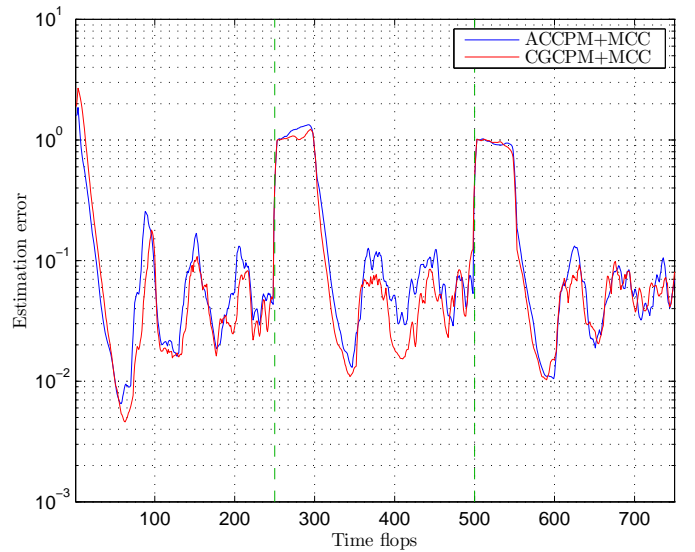}
\caption{Interference channel gain vector estimation error progress vs time of CPM methods using MCC feedback for slow fading channels}
\label{fig8}
\end{figure}
In Fig. \ref{fig8}, initially it has to be pointed that the learning error diagrams show variations, because the learning approach in the dynamic channel scenario is window based and not maximum likelihood based like in \cite{biban73}. Thus, the results have peaks and valleys instead of being smooth. Nevertheless, the advantage gained with this approach is that the obsolescence and thus the credibility of each inequality is not dependent any more on the arbitrary probit model and on a forgetting factor whose value choice is impractical. Moreover, the length of the window can be easily distinguished in every channel change where there is a constant average error of almost $100\%$ for $50$ time flops. This is caused because the learning algorithm in order to completely "forget" any inequality pair about the previous interference channel vector and proceed to the next one, a number of time flops equal to the observation window is necessary. It can also be observed that between the two CPMs the CGCPM delivers marginally less estimation error with only in one case surpassing the $10\%$ error barrier.   

\begin{figure}[!h]
\centering
\includegraphics[scale=0.735, trim=41 10 0 10, clip]{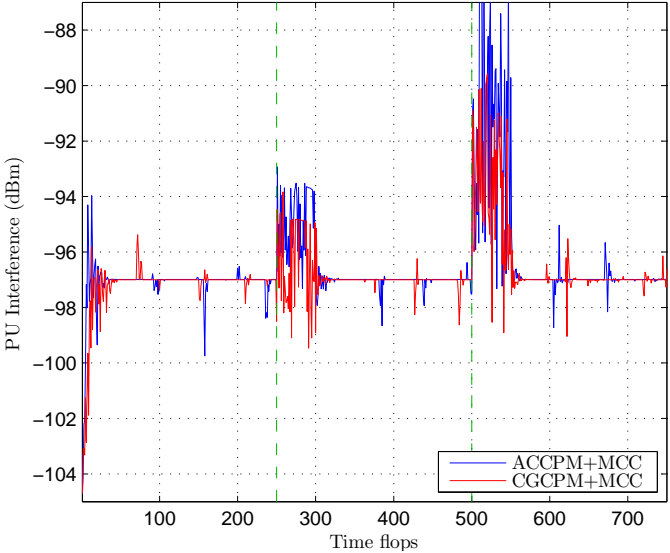}
\caption{$I_{PU}$ progress vs time using MCC feedback for slow fading channels}
\label{fig9}
\end{figure}

\begin{figure}[!h]
\centering
\includegraphics[scale=0.735, trim=40 10 0 10, clip]{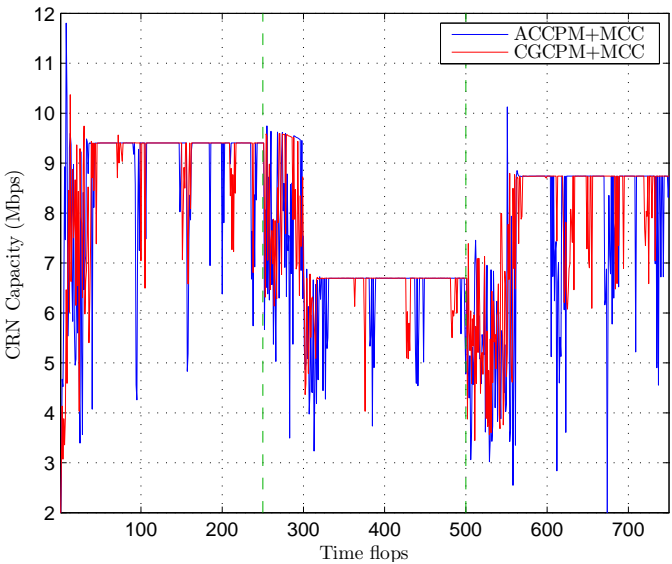}
\caption{CRN Capacity vs time using MCC feedback for slow fading channels}
\label{fig10}
\end{figure}

Next, we provide the aggregated interference and CRN capacity diagrams in Fig. \ref{fig9} and Fig. \ref{fig10} respectively for a single SU topology.

The main advantage observed in these diagrams of the CGCPM-based method over the ACCPM-based one is that despite the number of peaks and valleys which is roughly the same for both techniques, the CGCPM appears to have smaller variations in both diagrams. This provides better protection to the PU as shown in Fig. \ref{fig9}, since it causes less interference to the PU, and closer pursue of the optimization objective, the CRN capacity maximization, as shown in Fig. \ref{fig10}. Even though initially the main reason of using the CGCPM was its convergence rate, which theoretically is better, this advantage is not very clear in low dimensions and this can be seen in Fig. \ref{fig11}, where the convergence gain for N=5 SUs is only $5$ time flops and for N=10 SUs is increased to $10$ time flops in the MCC case.

In order to evaluate better the results of the diagrams in Fig. \ref{fig9} and Fig. \ref{fig10}, the average PU interference ($\olinea{I}_{PU}$) and the average CRN capacity ($\olineb{U}_{SU}^{tot}$) are calculated over the $3$ blocks for both methods and compared to derive further solid performance conclusions besides the convergence rate. For the ACCPM based method, these average metrics are $\olinea{I}_{PU}=-95.6\mbox{dBm}$ and $\olineb{U}_{SU}^{tot}=7.68\mbox{Mbps}$, while for the CGCPM based method they are $\olinea{I}_{PU}=-96.7\mbox{dBm}$ and $\olineb{U}_{SU}^{tot}=7.88\mbox{Mbps}$. We notice that the CPM used in this paper, the CGCPM, delivers on average $-1.1\mbox{dB}$ less PU interference and $2.6\%$ more CRN capacity compared to the ACCPM used in \cite{biban73}. Basically, our enhancement contributes to better adaptation and faster learning especially for large CRNs, closer pursue of the optimization objective and most importantly better protection of the PU.


\section{Conclusions}

In this paper, we proposed a simultaneous PC and interference channel learning algorithm using the MCC feedback. This sensing output is more informative than the binary ACK/NACK feedback and easier to obtain, since it does not require the implementation of an actual PU decoder on the SU sensing module. The proposed technique was applied in a CR scenario where a CRN with centralized structure access the frequency band of a PU operating under an ACM protocol and learns the unknown interference channels while maximizing its total capacity. Newly introduced Active Learning methods, the CPMs, were utilized for the design of the algorithm and compared to a benchmark learning method we previously developed in \cite{biban80}. The chosen CPMs were the ACCPM and the CGCPM inspired by the cognitive beamforming mechanism developed in \cite{biban73}. Additionally, a window-based solution was introduced for the case of slow fading interference channels. Initially, the results prove the superiority of the MCC feedback whose use provides us an implicit CSI of the PU link more informative than the binary feedback and thus delivers faster convergence. Subsequently, a comparison of the methods was performed which points out the better learning rate of the CPMs to the benchmark method and the small but yet distinguishable, especially in large CRNs, difference between the CGCPM-based approach and the ACCPM-based one. The CGCPM-based algorithm manages to be faster in static interference channel scenarios, more adaptive, more protective to the PU and with less variations in dynamic interference channel scenarios, due to its more intelligent choice of probing power vectors. An extension of this work could be the probabilistic version of the proposed algorithm which takes into account how accurate the output of the MCC process is by utilizing a reliability factor for each feedback. Even though this issue was addressed using a maximum likelihood approach in \cite{biban73}, still the proposed solution was not consistent in the Active Learning framework.



\bibliographystyle{IEEEtran}
\bibliography{Bibliografia}


\end{document}